\documentclass[12pt]{article}
\usepackage[margin=.75in]{geometry}
\usepackage{amsmath, amssymb, amsthm}
\usepackage{mathtools}
\usepackage{graphicx}
\usepackage{array}
\usepackage{tikz}
\usepackage{amsfonts}

\pdfoutput=1
\usepackage{putex}
\usepackage[vcentermath]{youngtab}
\usepackage{subfig}
\usepackage{lscape}
\usepackage{cite}
\usepackage{graphicx}
\usepackage{epstopdf}
\usepackage{enumerate}
\usepackage{tensor}
\usepackage{slashed}
\usepackage{amsmath}
\usepackage{amssymb}
\usepackage{mathrsfs}
\usepackage{lgrind}
\newcommand{\N}{\mathcal{N}}

\usepackage{amsmath,amssymb,braket}

\usepackage{bbm}

\usepackage{hyperref}

\numberwithin{equation}{section}

\newcommand {\be} {\begin {equation}}
\newcommand {\ee} {\end {equation}}

\newcommand {\bes} {\begin {equation*}}
\newcommand {\ees} {\end {equation*}}

\newcommand{\eps}{\epsilon}

\def\CH{{\cal H}}

\newcommand{\beq}{\begin{equation}}
\newcommand{\eeq}{\end{equation}}

\def\be{ \begin{equation} }
\def\ee{ \end{equation} }

\def \be {\beta}

\def \beq { \begin{equation}}
\def \eeq {\end{equation}}

\usepackage{indentfirst}
\begin{document}
\preprint{PUPT-2594}
	
\institution{PU}{Department of Physics, Princeton University, Princeton, NJ 08544}
\authors{Fedor~K.~Popov}
\title{Supersymmetric Tensor Model\\ at Large $N$ and Small $\epsilon$} 
\abstract{We study the $O(N)^3$ supersymmetric quantum field theory of a scalar superfield $\Phi_{abc}$ with a tetrahedral interaction. In the large $N$ limit the theory is dominated by the melonic diagrams. We solve the corresponding Dyson-Schwinger equations in continuous dimensions below $3$. 
For sufficiently large $N$ we find an IR stable fixed point and computed the $3-\epsilon$ expansion up to the second order of perturbation theory, which is in agreement with the solution of DS equations. We also describe the $1+\eps$ expansion of the model and discuss the possiblity of adding the Chern-Simons action to gauge the supersymmetric model.}
	
\date{}
\maketitle
\tableofcontents
\section{Introduction and summary}
In recent literature, there has been strong interest in theories whose dynamical fields are tensors of rank $3$ or higher (for reviews, see \cite{ Gurau:2016cjo, Rivasseau:2016zco,Klebanov:2018fzb}). Such theories possess a number of interesting features. For example, only the melonic diagrams dominate in the large $N$ limit, in contrast to the vector models, where only snail diagrams dominate \cite{Klebanov:2018fzb}, and the matrix models, where all the planar diagrams survive in the large $N$ limit. This fact makes the tensor models similar to the famous Sachdev-Ye-Kitaev (SYK) model \cite{Sachdev:1992fk,Kitaev:2015,Maldacena:2016hyu}. 
The SYK model contains a disordered coupling constant, making it hard to use standard tools of quantum field theory. The SYK model is believed to describe quantum properties of the extremal charged black holes \cite{Maldacena:2016upp,Engelsoy:2016xyb,Jensen:2016pah} and therefore may help to serve as a toy model for understanding the AdS/CFT correspondence \cite{Maldacena:1997re,Gubser:1998bc, Witten:1998qj}. It is already used for understanding the properties of the traversable wormholes \cite{Gao:2016bin, Maldacena:2017axo,Maldacena:2018lmt,Kim:2019upg}. While the tensor models \cite{Gurau:2016cjo} exhibit the same properties at the large $N$ limit, they do not have disorder therefore giving us hope that they can be understood at finite $N$ via standard techniques of quantum field theories. These techniques have already brought many interesting results \cite{Minwalla:2017,Mironov:2017aqv,Gurau:2011aq,Witten:2016iux,Klebanov:2016xxf,Prakash:2017hwq,Azeyanagi:2017mre,Benedetti:2017qxl,Benedetti:2018goh,
Itoyama:2017xid,Itoyama:2017wjb}.

We shall consider a supersymmetric analogue of such theories, which has been recently considered as a generalization of SYK model \cite{Fu:2016vas,Murugan:2017eto,Bulycheva:2018qcp} or as a quantum mechanical supersymmetric tensor model \cite{Chang:2018sve,Peng:2016mxj,Peng:2017spg, Chang:2019yug}. Here we will present a similar model in continuous dimension $d$. We consider a minimal $\mathcal{N}=1$ supersymmetric model, where we have some number of scalar superfields $\Phi_{abc}(x,\theta)$, and indices $a,b,c$ run from 1 to $N$. These fields are coupled via a ``tetrahedral`` superpotential\footnote{Here we will refer to the appendix \ref{supersection} and the paper \cite{gates2001superspace} for the notations and the other helpful formulas that will be used through the paper.}
\begin{gather} \label{smelon}
S = \int d^d x\, d^2 \theta\,\left[\frac12 \left(D_\alpha \Phi_{abc}\right)^2 + g \Phi_{abc}\Phi_{ab'c'}\Phi_{a'b c'} \Phi_{ab'c'}\right] .
\end{gather}
This theory, which is renormalizable in $d<3$,  possesses $O(N)\times O(N) \times O(N)$ symmetry rather than $O(N^3)$ (the superpotential breaks such a symmetry, while the free theory, of course, posses the $O(N^3)$ symmetry). This model has been proposed in the paper \cite{Klebanov:2016xxf} as a generalization of the scalar melonic theory. It was proved that the non-supersymmetric analogue of this theory has a so-called melonic dominance in the limit when $N\to \infty, g \to 0$ but $g N^\frac32$ is kept fixed \cite{Carrozza:2015adg}. The proof of this peculiar fact relies on the combinatorial properties of the potential, and therefore is applicable in any dimensions and in various theories, provided that the combinatorial properties are left the same. In the case of the supersymmetric theories, the Feynman diagrams, written down in terms of the components, look quite complicated and, at first glance, do not possesses a melonic limit as in the case of scalar model or the SYK model. However, one can develop a supersymmetric version of the usual Feynman diagrams technique and work explicitly with the superfields $\Phi_{abc}$ and see that the combinatorial and topological properties are the same as in the case of the scalar tensor models. Therefore, the proof of the dominance of melonic diagrams \cite{Witten:2016iux, Pallegar, Klebanov:2016xxf,Carrozza:2015adg, Carrozza:2018ewt} is applicable in this case and the theory \eqref{smelon} also possesses a melonic dominance in the large $N$ limit. We generalize the theory \eqref{smelon} where the tethrahydral term is replaced by $q$-valent maximally single-trace operator to study models with different numbers of the internal propagators in each melon \cite{Ferrari:2017jgw,Pallegar}.

The properties of such theories in the IR limit can be investigated by solving the Dyson-Schwinger (DS) equations, which are drastically simplified if the theory is melonic. Namely, the dominance of the melonic diagrams in the large $N$ limit can be understood as a suppression of the corrections to the vertex operators in the system of DS equations.
The solution of the DS equation in the IR yields a conformal propagator, suggesting that the theory in the IR flows to the fixed point, which is described by some conformal field theory. The existence of the stress-energy tensor with the correct dimension, and the spectra of the operators confirm this hypothesis. Therefore, one can wonder whether it is possible to describe such a transition from the UV scale (where we have a bare conformal propagator determined by commutation relations) to the IR region by means of RG flow and $\eps$ expansion. Several attempts have been made towards this idea. For example, the melonic scalar theory in $4$ dimensions \cite{Giombi:2017dtl} has been considered at the second order of the perturbation theory. For this theory, a melonic fixed point of RG flow was found, even though the corresponding couplings are complex.  The complex couplings indicate that the theory is unstable. For example, the dimensions of some operators have imaginary part. One of the reasons of instability could be that the potential is unbounded from below, leading to the decay of the vacuum state. The theory \eqref{smelon}, being supersymmetric, lacks such a disadvantage. 

It is quite interesting that if one drops the fermionic part of the action \eqref{smelon} and integrates out the auxiliary field, the theory still possesses the melonic dominance in the large $N$ limit.  Such a "prismatic" theory was considered in the paper \cite{Giombi:2018qgp}. The solution of this theory was found in the large $N$ limit and the RG properties were investigated at two loops. As opposed to the standard melonic theory \cite{Giombi:2017dtl}, the fixed point is real and first order of $\epsilon$ expansion recovers the exact solution in the large $N$ limit. 

In this paper we solve the model \eqref{smelon} in the large $N$ limit, assuming that the supersymmetry is not broken and that in the IR region the theory is described by the conformal propagator. The solution is found for general dimension $d$ and general $q$-valent MST potential \cite{Ferrari:2017jgw,Pallegar}. The dimension of the operators at given $d$ and spin $s$ can be found as a solution of the corresponding transcendental equation. It is shown that at any dimension $d$, there is always a stress-energy operator of dimension $d$ and a supercurrent operator of dimension $d-\frac12$, which indicates that the theory is indeed described by a conformal field theory. While the model \eqref{smelon} exists only in the fractional dimensions between one and three dimensions, the counterpart SYK model with $q=3$ can work at the integer dimension $d=3$ and describe a good conformal field theory with the melonic dominance in the large $N$ limit. After that we derive a perturbation theory in $3-\eps$ dimensions  of the theory \eqref{smelon} to find a fixed point that could describe the IR solution of the large $N$ limit of the model \eqref{smelon}. We find that the $\epsilon$ expansion is consistent with the exact large $N$ solution up to the first order in $\epsilon$. The two-loop analysis also suggests that the found melonic fixed point is IR stable. 

The structure of the paper is as follows: in section two, we discuss the properties of the theory \eqref{smelon} in the large $N$ limit. The dimensions of the operators are found and the DS equation is solved in the superspace formalism. In section three, we consider $q=3$ supersymmetric SYK model and study the stability of such a theory. In section four, we study the RG properties of the quartic super theories in 3 dimensions and compare to the exact solutions in the large $N$ limit.  In section five, we discuss the possibility of introducing higher order supersymmetry and speculate about the consequences of gauging the supersymmetric tensor models. The appendix provides  supplemental materials including the notations and useful formulas that are used throughout the paper.

\section{Solution of the Large $N$ Theory}

\begin{figure}
	\centering
	\includegraphics[scale=1]{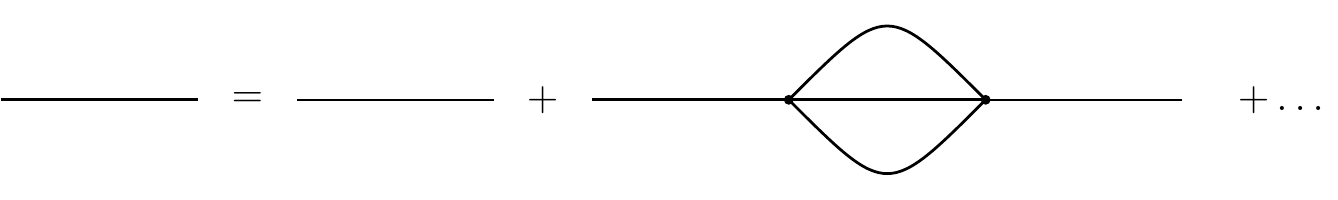}
%
%
%
%
%
	\caption{A supersymmetric version of the Dyson-Schwinger equation for melonic theories in the large N limit.}
	\label{superDSfig}
\end{figure}

In this section, we will try to find the solution of DS equations for the theory \eqref{smelon} in the large $N$ limit. As mentioned in the introduction, the theory possesses a melonic dominance in the large $N$ limit. This means that only specific diagrams survive in the large $N$ limit, namely the ones generated recursively by the Dyson-Schwinger (DS) equation (schematically depicted in the fig.\eqref{superDSfig}). The resulting equation for scalar or fermion field theories was investigated analytically and numerically for many different theories \cite{Maldacena:2016hyu,Klebanov:2016xxf,Patash:1964sp}. For example, the DS equation can be solved in the IR limit and the solution possesses a conformal symmetry in that limit. In the case of the supersymmetric theories, one of the important differences is that one can demand the solution to respect supersymmetry. In order to do it manifestly the DS equation should be formulated in terms of the superfields. Of course, one can do this calculation in terms of the components as in the paper \cite{Chang:2018sve} and check that these two approaches give the same answers. To make the discussion more general we consider the case where there are $q-1$ internal propagators in the melon diagrams and suitable MST operator is considered \cite{Pallegar}. The DS equation in the supersymmetric case reads as
\begin{gather}\label{DSeq}
G(p; \theta, \theta')= G_0(p;\theta,\theta') + \\+ \frac{1}{16}\lambda^2\int d^2\theta_1\, d^2 \theta_2\, G_0(p;\theta,\theta_1) \int \prod^{q-1}_{i=1}\frac{d^d k_i}{(2\pi)^{d}} G(k_i;\theta_1,\theta_2) (2\pi)^d \delta^d\left(p-\sum^{q-1}_{i=1} k_i\right)G(p;\theta_2,\theta') \notag,
\end{gather}
where $G_0(p;\theta,\theta')$ is a bare superpropagator \eqref{sprop}, $G(p;\theta,\theta')$ is an exact superpropagator and $g =\lambda N^\frac32$ is a 't Hooft coupling. Analogously to the scalar case, we consider a conformal propagator as an ansatz for the solution. But if we also demand to preserve supersymmetry and $O(N)\times O(N) \times O(N)$ symmetry, that yields only one form of the solution
\begin{gather}\label{sans}
\braket{\Phi_{abc}(p,\theta) \Phi_{a'b'c'}(-p,\theta')} = \delta_{aa'}\delta_{bb'}\delta_{cc'} G(p;\theta,\theta'), \quad G(p;\theta,\theta') = A \frac{D^2 \delta(\theta-\theta')}{p^{2\Delta}},
\end{gather}
where $\Delta < \Delta_0 = 1$ for the solution to be valid in the IR limit \cite{Maldacena:2016upp} (namely, we can neglect by bare propagator in comparison to the exact one $G_0^{-1} \ll G^{-1}, p \to 0$ ). Substituting the ansatz in the DS equation \eqref{DSeq} we get
\begin{gather}
 A \frac{D^2 \delta(\theta-\theta')}{p^{2\Delta}}=   \frac{D^2 \delta(\theta-\theta')}{p^{2}} + \\+ A^q\lambda^2\int d^2\theta_1\, d^2 \theta_2\,  \frac{D^2 \delta(\theta-\theta_1)}{p^2} \prod^{q-1}_{i=1}\int \frac{d^d k_i}{(2\pi)^{d}}   (2\pi)^d \delta^d\left(p-\sum^{q-1}_{i=1} k_i\right) \frac{D^2\delta(\theta_1-\theta_2)}{k_i^{2\Delta}}\frac{D^2 \delta(\theta_2-\theta')}{p^{2\Delta}}. \notag
\end{gather}
As soon as $\Delta<1$ we can neglect the LHS of the equation by the RHS in the limit $p\to 0$. After that one can integrate out Grassman variables using  identities for the superderivative to get
\begin{gather}
\lambda^2 A^q \prod^{q-1}_{i=1}\int \frac{d^d k_i}{(2\pi)^{d}} \frac{1}{k_i^{2\Delta}}  (2\pi)^d \delta^d\left(p-\sum^{q-1}_{i=1} k_i\right)  \frac{1}{p^{2\Delta-2}} =- 1.
\end{gather}
This equation gives the dimension of the superfield to be $\Delta = \frac{d(q-2)+2}{2 q}$ and
\begin{gather}
A^q = \frac{(4\pi)^{\frac{d(q-2)}{2}}}{\lambda^2} \frac{\Gamma^{q-1}\left(\frac{d}{2}-\frac{d-1}{q}\right)\Gamma\left(d-1-\frac{d-1}{q}\right)}{\Gamma^{q-1}\left(\frac{d-1}{q} \right)\Gamma\left(\frac{d-1}{q}-\frac{d}{2}+1\right) }.
\end{gather}
The solution suggests that we cannot work directly in $d_{\rm crit}(q)=\frac{2 q -2}{q-2}$ dimensions because the bare propagator is not suppressed in the IR limit and change the solution. For example, for the case of tetrahydral potential $q=4$, $d_{\rm crit}=3$, therefore the tensorial melonic theory is not conformal in 3 dimensions.
Nevertheless, we can still study the theory slightly below 3 dimensions and compare it with the $\eps$ expansion. 

If one chooses the case of $q=3$, the critical dimension is $d_{\rm crit} = 4$ and such a melonic theory should describe a conformal field theory in 3 dimensions. In the next section we will review this model in more details.

We calculated the propagator \eqref{sans} in the momentum representation. One can carry out the calculation in the coordinate space. With the use of the relation
\begin{gather}
\int \frac{d^d k}{(2\pi)^d}  e^{i k x }D^2 \delta(\theta-\theta')
=\int \frac{d^d k}{(2\pi)^d} e^{i k x}\left(1-i k^\mu \bar{\theta'}\gamma_\mu\theta + k^2 \bar{\theta'}\theta' \bar{\theta}\theta\right) = e^{\bar{\theta'} \gamma^\mu \theta \frac{\partial}{\partial x^\mu}},
\end{gather}
the propagator in the coordinate representation is
\begin{gather}
G(x,\theta,\theta')= \frac{B}{|x_\mu - \bar{\theta'}\gamma_\mu \theta|^{\frac{2(d-1)}{q}}},\quad B^q = \frac{1}{4 \pi^d\lambda^2} \frac{\Gamma\left(\frac{d-1}{q} \right)\Gamma\left(d-1-\frac{d-1}{q}\right)}{\Gamma\left(\frac{d}{2}-\frac{d-1}{q}\right)\Gamma\left(\frac{d-1}{q}-\frac{d}{2}+1\right) }.
\end{gather}
Another way to see that the dimension of the superfield is $\frac{d-1}{q}$ is to rewrite the action in terms of the components and impose the conditions $\Delta_\psi = \Delta_\phi+\frac12$, then the action contains a term
\begin{gather}
W(\Phi) = \Phi^q \Rightarrow W(\phi) = \phi^{q-2} \psi^2 \Rightarrow [W]=d\Rightarrow (q-2)\Delta_\phi + 2 \Delta_\psi = d,\quad \Delta_\phi = \frac{d-1}{q}.
\end{gather} 
The solution \eqref{sans} suggests that in the IR limit, the theory is described by some conformal field theory (CFT). One of the interesting questions that one may ask is, what is the spectrum of the bipartite conformal operators in this theory? 
The supersymmetric theory \eqref{smelon} has different types of the bipartite operators, as the prismatic one \cite{Giombi:2018qgp}. We should consider these families separetly. The most simple ones have the following structure \cite{Murugan:2017eto}
\begin{gather} \label{sprim}
V_{FF} = \Phi_{abc}(x,\theta) \Box^h \Phi_{abc}(x,\theta), \quad V_{BB} = \Phi_{abc}(x,\theta) \Box^h D^2 \Phi_{abc}(x,\theta). 
\end{gather}
These operator should be considered as a collection of operators with different spins and dimensions, that transforms through each other when the supersymmetry transformations are applied. For shorthand, we will omit the indecies, assuming that the operators are singlet under the action of $O(N)$'s groups. These operators could be rewritten in the terms of components \eqref{superfield} as
\begin{gather}
V_{FF} (x,\theta) = \phi(x)\Box^h \phi(x) + \phi(x) \Box^h \psi^\alpha(x) \theta_\alpha + \theta^2\left(\phi(x) \Box^h F(x) + \Box^h\phi(x)  F(x)+ \bar{\psi}(x)\Box^h \psi(x)\right),\notag\\
V_{BB}(x,\theta) =\bar{\psi}\Box^h\psi+\left(F \Box^h\psi_\alpha + \Box^h F \psi_\alpha + (\gamma^\mu\psi)_\alpha \partial_\mu \Box^h \phi + (\gamma^\mu \Box^h\psi)_\alpha \partial_\mu \Box^h \phi \right)\theta^\alpha+ \notag\\+\theta^2\left(\partial_\mu \phi \Box^h \partial_{\mu}\phi + i \bar{\psi}\gamma^\mu \Box^h\partial_{\mu} \psi + F\Box^h F\right).
\end{gather} 
A similar set of the operators was considered in the paper \cite{Murugan:2017eto} in 2 dimensions and \cite{Chang:2018sve} in 1 dimension. Later we shall compare the results of these papers with the continuous solution for arbitrary $d$. We can try to put more $D^2$ in \eqref{sprim} to get more familes, but with the use of the identity $(D^2)^2=\Box$, one can descend these operators to the BB or FF series. That's why we can consider only these two families to get the whole spectrum of bipartite operators with the lowest component having spin $s=0$.

As usual, the corrections to the bilinear operators in the large $N$ limit are given by the ladder diagrams (but again, in comparison to \cite{Maldacena:2016upp,Klebanov:2016xxf}, these diagrams should be considered to be in superspace). We assume the following ansatz in momentum space for the three-point correlation function for these families,
\begin{figure}
	\centering
%
%
	\includegraphics{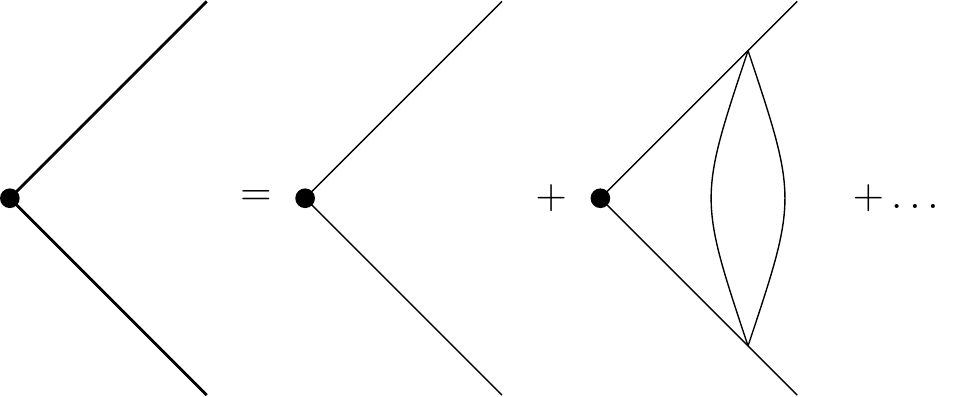}
	\caption{The corrections to the bipartite conformal operator can be summed with the use of the Bethe-Salpeter equation. The diagrams should be considered to be in the superspace.}
	\label{ladders}
\end{figure}
\begin{gather}
G_{FF}(k,\theta,\theta') = \braket{V_{FF}\Phi(-k,\theta) \Phi(k,\theta')} = \frac{\delta(\theta-\theta')}{k^{\Delta_V+2\Delta}}, \quad \notag\\
G_{BB}(k,\theta,\theta')=\braket{V_{BB}\Phi(-k,\theta) \Phi(k,\theta')} = \frac{D^2\delta(\theta-\theta')}{k^{\Delta_V+2\Delta}},
\end{gather}
where we have set the operators $V_{BB},V_{FF}$ to be at infinity and made a Fourier transformation with respect to the spatial coordinates, and $\Delta_V$ is the corresponding dimensions of the operator. The derivation of the equations for the dimensions $\Delta_V$ is just a straightforward generalization of the analogous calculation for the scalar model \cite{Klebanov:2016xxf} or the SYK model \cite{Maldacena:2016hyu}.  Here we will show the derivation of such equation for the $BB$ operators.

The addition of one step of the ladder can be considered as the action of the {\it kernel} operator,
\begin{gather}
\hat{K} = K(p,k;\theta,\theta',\theta_1,\theta_2) = \notag\\ = (q-1) \prod^{q-2}_{i=1}\int \frac{d^d q_i}{(2\pi)^d} \frac{D^2 \delta(\theta_1-\theta_2)}{q_i^{2\Delta}}  \frac{D^2 \delta(\theta-\theta_1)}{p^{2\Delta}}  \frac{D^2 \delta(\theta_2-\theta')}{p^{2\Delta}} (2\pi)^d \delta^{d}\left(\sum q_i - (p-k)\right). 
\end{gather}
We act on the \eqref{sprim} by one step of the ladder,
\begin{gather}
(\hat{K}G_{BB})(p,\theta,\theta')=\int d^2 \theta_1 d^2\theta_2 \int \frac{d^d k}{(2\pi)^d} K(p,k;\theta,\theta',\theta_1,\theta_2) G_{BB}(k,\theta_1,\theta_2). 
\end{gather} 
The Grassman variables can be integrated out with the use of identities from the section \ref{supersection}. After that we are left with a simple integral
\begin{gather}
(\hat{K}G_{BB})(p,\theta,\theta')= \notag\\= (q-1)A^q \lambda^2 D^2 \delta(\theta-\theta')\int \frac{d^d k}{(2\pi)^{d}} \prod^{q-2}_{i=1}\frac{d^d q_i}{(2\pi)^d} \frac{1}{q_i^{2\Delta}} \frac{1}{k^{\Delta_V + 2\Delta} p^{2\Delta-2}}  (2\pi)^d \delta^{d}\left(\sum q_i - (p-k)\right) = \notag\\
=g_B\left(\Delta_V\right) G_{BB}(p,\theta,\theta'),
\end{gather}
where
\begin{gather}
g_B(\Delta_V) = -(q-1)\frac{\Gamma\left(\frac{2+d(q-2)}{2 q}\right)\Gamma\left(\frac{(q-1)(d-1)}{4}\right) \Gamma\left(\frac{d}{4}-\frac{1}{q}-\frac{\Delta_V}{2}\right)\Gamma\left(1-\frac{d}{2}-\frac{1}{q}+\frac{d}{q}+\frac{\Delta_V}{2}\right)}{\Gamma\left(1-\frac{d}{2}+\frac{d-1}{q}\right)\Gamma\left(\frac{d-1}{q}\right)\Gamma\left(\frac{(q-1)(d-1)}{q}-\frac{\Delta_V}{2}\right)\Gamma\left(\frac{d}{2}+\frac{1}{q}-\frac{d}{q} + \frac{\Delta_V}{2}\right)}.
\end{gather}
In order for the operator to be primary, the equation $g_B(\Delta_V)=1$ must hold. An analogous equation can be written for the $V_{FF}$ operator, but one can see that
\begin{gather} \label{FBrelation}
g_F(\Delta_V) = g_B\left(\Delta_V-1\right),
\end{gather}
This suggests that we might build a bigger multiplet and enhance the supersymmetry to be $\N=2$ (later we shall see that this does not actually happen, because there is no additional fermionic counterparts to finish supermultiplet).

\begin{figure}
	\centering
	\includegraphics[scale=0.35]{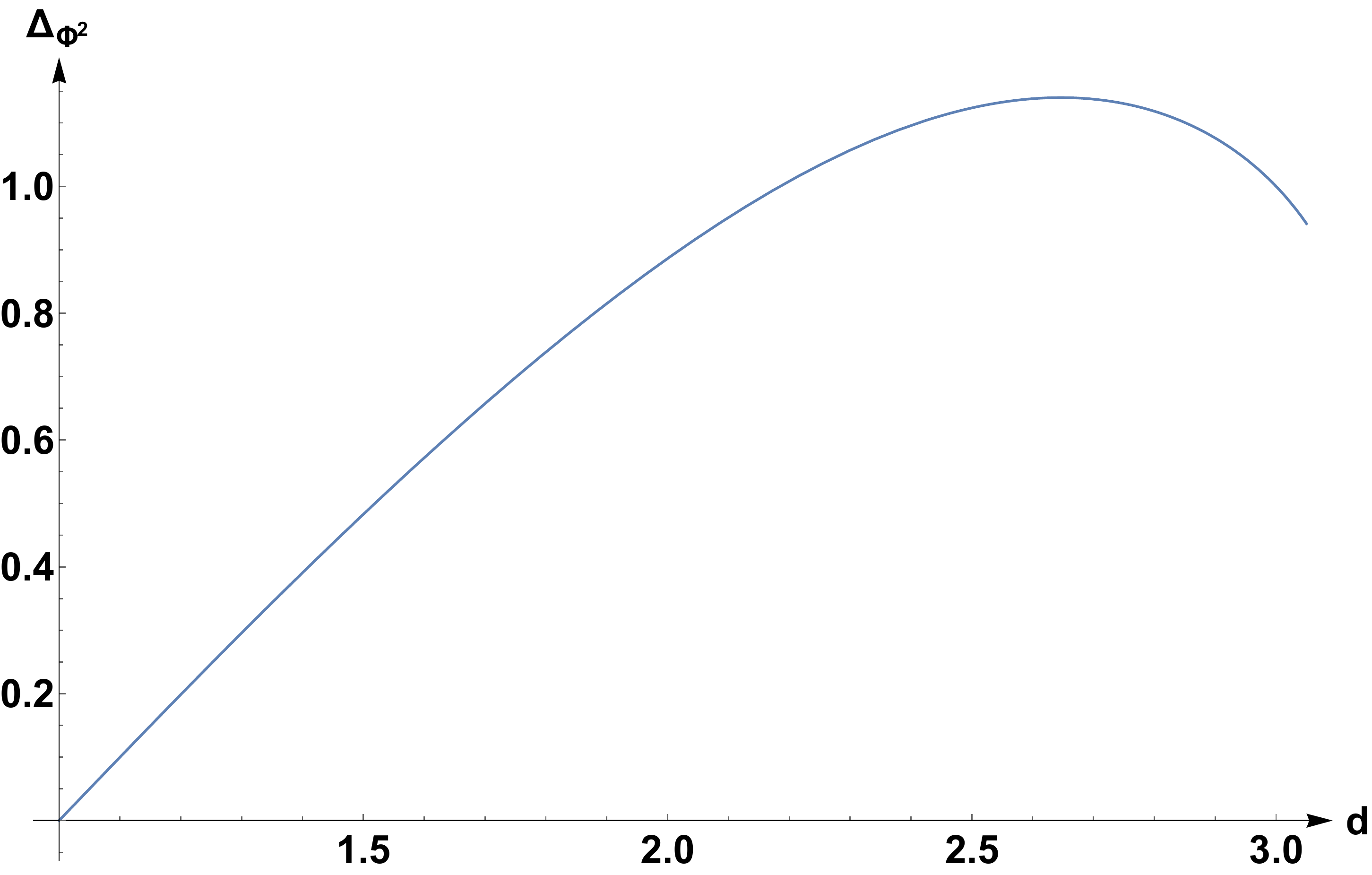}
	\caption{The dimension of the operator $\Phi^2$ as a function of the dimension. As $d\to 1$ the dimension goes to zero.}
	\label{fig:thedimphi2}
\end{figure}

From now on we shall consider the case only $q=4$ to get $3-\eps$ expansion unless the other is specified. Thus, we can get the $\eps$ expansion in the large $N$ limit of the $\Phi^2$ operator
\begin{gather}
\Delta_{\Phi^2} =  1 + \eps +3 \eps^2 - \frac{\pi^2+24}{4}\eps^3 + \mathcal{O}(\eps^4). \label{dimPhi2}
\end{gather}
The plot of the $\Delta_{\Phi^2}$ as a function of the dimension is depicted in the figure \ref{fig:thedimphi2}.
Analogously we get the dimension of $\Phi D^2 \Phi$ operator
\begin{gather}
\Delta_{\Phi D^2 \Phi} = 2 + \eps +3 \eps^2 - \frac{\pi^2+24}{4}\eps^3 + \mathcal{O}(\eps^4).
\end{gather}

We can discuss dimensions of non-singlet operators of the form $\Phi_{abc}\Phi_{a'bc}$. The equation for the dimension of this operator can be rewritten as
\begin{gather}
g_B(\Delta_{aa'}) = q-1,
\end{gather}
where a factor $q-1$ appears from the combinatorics \cite{Bulycheva:2017ilt}, and $\Delta_{aa'}$ is the dimension of the operator. The $\eps$ expansion near three dimensions for $q=4$ has the following form
\begin{gather}
 \Delta_{aa'} = 1 - \frac12 \eps^2 + \frac{\pi^2}{24}\eps^3+ \mathcal{O}(\eps^4).
\end{gather}
Later, we shall show that the solution coincides with the $\eps$ expansion in the second level of perturbation theory.

From this, the next step would be to study the spectrum of the higher-spin operators. A generalization for the higher spin operators is
\begin{gather}
V_{FF}^s = \Phi(x,\theta) \Box \partial_{\mu_1}\ldots \partial_{\mu_s} \Phi(x,\theta), \quad V_{BB}^s=\Phi(x,\theta) \Box \partial_{\mu_1}\ldots \partial_{\mu_s} D^2 \Phi(x,\theta),
\end{gather}
with the corresponding modifications for the ansatz. For example, for higher spin spectrum of the BB operators the ansatz is
\begin{gather}
G^s_{\mu_1\ldots \mu_s,BB}(k,\theta,\theta')=\braket{V^s_{\mu_1\mu_2\ldots \mu_s,BB}\Phi(-k,\theta) \Phi(k,\theta')} = \frac{D^2\delta(\theta-\theta') k_{\mu_1}\ldots k_{\mu_s}}{k^{\Delta_V+\frac{d+1}{2}+s}}. \label{spinansatz}
\end{gather}
In this case we consider an arbitrarily chosen null-vector $\xi^\mu$ and consider the convolution of the ansatz \eqref{spinansatz} with the vector $\xi$. After that one can integrate out the Grassman variables and carry out the integration over the real pace with the use of a relation \cite{Giombi:2017dtl}:
\begin{gather}
\int d^d x \frac{(\xi \cdot x)^s}{x^{2\alpha} (x-y)^{2\beta}} = \pi^\frac{d}{2} \frac{\Gamma\left(\frac{d}{2}-\alpha+s\right)\Gamma\left(\frac{d}{2}-\beta\right)\Gamma\left(\alpha+\beta-\frac{d}{2}\right)}{\Gamma(\alpha)\Gamma(\beta)\Gamma(d+s-\alpha-\beta)} \frac{(\xi \cdot y)^s}{y^{2\alpha+2\beta-d}}.
\end{gather}
\begin{figure}
	\centering
	\includegraphics[scale=0.55]{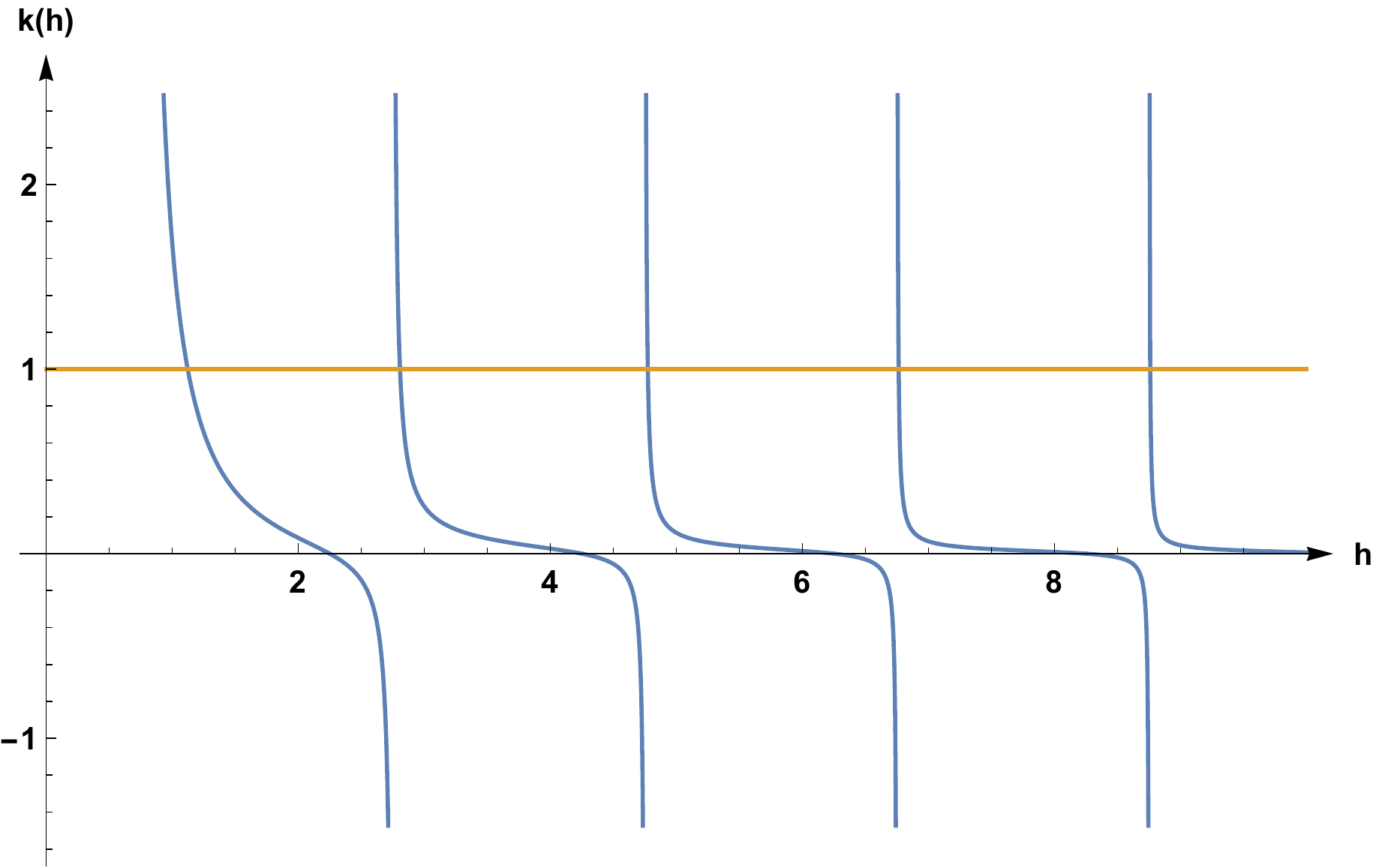}
	\caption{The dimension of the operator $\Phi^2$ can be found graphically. The plot of $k(h)$ is drawn for the case of $d=2.5$}
	\label{fig:kbb25}
\end{figure}

Eventually, the equation for the dimension at given spin $s$ reads as
\begin{gather}\label{spinBFF}
g_B(d,\Delta_V,s) =\\= -(q-1)\frac{\Gamma\left(\frac{2+d(q-2)}{2 q}\right)\Gamma\left(\frac{(q-1)(d-1)}{4}\right) \Gamma\left(\frac{d}{4}-\frac{1}{q}-\frac{\Delta_V-s}{2}\right)\Gamma\left(1-\frac{d}{2}-\frac{1}{q}+\frac{d}{q}+\frac{\Delta_V+s}{2}\right)}{\Gamma\left(1-\frac{d}{2}+\frac{d-1}{q}\right)\Gamma\left(\frac{d-1}{q}\right)\Gamma\left(\frac{(q-1)(d-1)}{q}-\frac{\Delta_V-s}{2}\right)\Gamma\left(\frac{d}{2}+\frac{1}{q}-\frac{d}{q} + \frac{\Delta_V+s}{2}\right)}=1,\notag
\end{gather}
One would expect that there is a solution at any $d$ and $s=2$ with $\Delta=d$, because of the existence of the stress-energy tensor. However one cannot find this solution. The reason is quite simple. First of all, there is no stress-energy tensor in the field decomposition of the BB and FF operators. Second, the stress-energy tensor has a superpartner $\mathcal{S}_\mu^\alpha$ (corresponding to supertranslations) that has spin $\frac32$, and therefore to find it we should consider a whole different family of operators, with lowest component being a Rarita-Schwinger field . Namely, let us consider a Fermi conformal primary operator
\begin{gather}\label{BFoper}
V_{BF,\mu_1\ldots \mu_{2n+1}}(x,\theta) =  \partial_{\mu_i}^{2n+1}\Phi(x,\theta) D_\alpha\Phi(x,\theta),
\end{gather}
where the odd number of the space-time derivatives should be inserted to get a primary operator. Indeed, if we consider a zero number of the derivatives
\begin{gather}\label{BFcons}
V_{BF} = \Phi_{abc} D_\alpha \Phi_{abc} = \frac12 D_\alpha \left(\Phi_{abc}^2\right),
\end{gather}
it is just a descendant of the FF operator. To get a supercurrent multiplet we have to project the operators \eqref{BFoper} on the specific component.   The ansatz for the three-point function has the following form 
\begin{gather}
\braket{V_{BF}\Phi(k,\theta)\Phi(-k,\theta')} = \frac{D_\alpha\delta(\theta-\theta')}{k^{\Delta_V+2\Delta}}.
\end{gather}
The derivation of the equation for the spectrum of the dimensions is straightforward
\begin{gather}\label{BBFrelation}
g_{BF}(d,\Delta_V,s) =- g_{B}\left(d,\Delta_V-\frac12,s - \frac12\right)=1,
\end{gather} 
where the spin should be chosen to be of the form $s=2n-\frac12$.
Now we can try to find the stress-energy momentum and its partner. And indeed at any $d,q$ and $s=\frac32$ there is an operator with dimension $\Delta=d-\frac12$ that corresponds to the usual stress-energy supermultiplet.  

At this point one can wonder whether the current  $J_{aa'}$, responsible for the $O(N)$'s transformations, is a primary operator. The supersymmetric multiplet containing the current should be also a Fermi supermultiplet with spin $s=1/2$ (this operator is not a singlet operator and therefore \eqref{BFoper} is not applicable). The current should satisfy the equation \cite{Bulycheva:2017ilt}
\begin{gather}
g^{aa'}_{BF}(d,\Delta_V,s)  = \frac{1}{3} g_{BF}(d,\Delta_V,s)= 1,
\end{gather}
at any $d$ and $q$ there is always a solution $\Delta_V=d-3/2$.  One can see that the dimension of  square of this operator is given by the direct sum of the dimensions $\Delta_{J\tilde{J}}=2 \Delta_V =2d-3$. This operator becomes relevant when $\Delta_{J\tilde{J}} = 2 d-3 \leq d-1$, where minus 1 comes from the accounting the dimension of the superspace. From this one can see the operator becomes marginal in $d=2$ and relevant as $d<2$. This extra marginal operator in $d=2$ may destabilize the CFT. The only exception is the case $N=1$, where the theory does not have any continuous symmetry and has superpotential $\Phi^4$. In $d=2$ this theory flows to the $m=4$ superconformal minimal model, which has central charge
$c=1$. \footnote{I would like to thank I.R.Klebanov for pointing out these facts.}

The relation \eqref{spinBFF} can be thought as a generalization of the equation for the kernel at 2 dimensions derived by Murugan et al. \cite{Murugan:2017eto}. In this case they introduced two dimensions, $h=\frac{\Delta+s}{2}$ and $\tilde{h} = \frac{\Delta-s}{2}$, and one can check that
\begin{gather}
k(h,\tilde{h}) = g_B(d=2,h+\tilde{h},h-\tilde{h})  = -(q-1)\frac{\Gamma^2(1-1/q) \Gamma(1/q-\tilde{h})\Gamma(1/q+h)}{\Gamma^2(1/q)\Gamma(1-1/q-\tilde{h})\Gamma(h+1-1/q)},
\end{gather}
which coincides with the equation (7.17) in \cite{Murugan:2017eto}.

The relation \eqref{BBFrelation} also shows that if there is a scalar bilinear multiplet with dimension $h$, there is no $BF$ operator with higher spin and the dimension $\Delta=\Delta+\frac12$. This shows that we cannot complete the $\mathcal{N}=2$ supermultiplet and the enhancement does not happen. It is interesting that there is an argument in $d=1$ stating that it actually must happen. Basically, it comes from the fact that group of diffeormorphims of supertransformations in 1-dimension comprises the $\mathcal{N}=2$ superalgebra \cite{Murugan:2017eto}. 


Finally we discuss the dimension of the quartic operators, because there is a fundamental relation between their dimensions and the eigenvalues of the matrix $\frac{\partial \beta_i}{\partial g_j}$. We can find the dimensions of some quartic operators in the large $N$ limit. For example, in the matrix models the anomalous dimension of a double trace operator is just the sum of the anomalous dimensions of the corresponding single trace operators. 
By the same analysis, we get that the anomalous dimension of the double trace operator is
\begin{gather}
\Delta_{\Phi^4} = 2\Delta_{\Phi^2} = 2 + 2\eps +\mathcal{O}(\eps^2).
\end{gather}
Analogous analysis gives that
\begin{gather}
\Delta_{\rm Pillow} = 2 \Delta_{aa'} = 2 + \mathcal{O}(\eps^2).
\end{gather}
Finally, the dimension of the tetrahedral operator can be determined as the dimension of the operator $\Phi_{abc}D^2\Phi_{abc}$ (namely, it follows from the equations of motion) and it gives us
\begin{gather}
\Delta_{\rm Tetra} = 2 + \eps +\mathcal{O}(\eps^2).
\end{gather} 
\subsection{The $\eps$ expansion near one dimension}
One can try to study the behaviour of the model \eqref{smelon} near $1$ dimension. The case of $d=1$ supersymmetric tensor models was considered recently (see \cite{Chang:2018sve}). It was found that the supersymmetry is broken in the IR region. The easiest way to see this is to assume a conformal ansatz and plug it in the DS equation \eqref{DSeq}. The solution suggests $\Delta=0$ in one dimensions, but constant or logarithm function do not satisfy the DS equation. The conformal solution found in \cite{Chang:2018sve} shows, that the dimensions of the superfield components are not related to each other by usual supersymmetric relations. It might be the case that for the system in 1 dimensions the conformal solution does not describe the true vacuum state, while the true vacuum 
respect supersymmetry and the propagators exponentially decays at large distances. It might be shown by studying the stability of the conformal solution in a way described in \cite{Kim:2019upg} for two coupled SYK models.
 
Also, if one consider a limit $d\to 1$ in the equations derived in the previous sections, the propagator does not have a smooth limit in 1 dimension and the kernel is equal to the constant $\lim\limits_{d\to 1} g_B(d,h,s) = -1$. The last fact confirms that in 1 dimension the conformal IR solution does not respect the supersymmetry. But, in the vicinity of dimension 1, everything works fine. Thus, one can study the $1+\eps$ expansion.  We shall consider the case of tensor models and set $q=4$. For example, the dimension of the $\Phi^2$ operator is
\begin{gather}
\Delta_{\Phi^2} = \eps-\frac{\pi^2}{48} \eps^3+\frac{3\zeta(3)}{16}\eps^4+\mathcal{O}(\eps^5),\quad \Delta_{\Phi D^2 \Phi}=1+ \eps-\frac{\pi^2}{48} \eps^3+\frac{3\zeta(3)}{16}\eps^4+\mathcal{O}(\eps^5).
\end{gather}
And the dimension of the colored operators $\Phi_{abc}\Phi_{a'bc}$ is
\begin{gather}
\Delta_{aa'} = \frac{3}{4}\eps - \frac{3\pi^2}{256}\eps^3+\frac{9 \zeta(3)}{128} \eps^4+\mathcal{O}(\eps^5)
\end{gather}
It would be interesting to derive this results by considering a one dimensional supersymmetric melonic quantum mechanics and lift the solution to $1+\eps$ dimenion. Or just derive these results starting with the conformal solution found in one dimension \cite{Chang:2018sve} and show that in higher dimensions the supersymmetry is immediately restored.

\section{Supersymmetric SYK model with $q=3$ in $d=3$}
In the previous section we mostly work with the tensor models in non-integer dimensions. The main problem that did not allow us to work directly in 3 dimensions was that the critical dimension for such a interaction is $d_{\rm cr} = \frac{2 q -2}{q-2}=3$, meaning that directly at 3 dimensions the conformal IR solution does not work. Nevertheless, if one considers $q=3$ case the critical dimension becomes $d_{\rm cr} = 4$ and therefore should work perfectly in 3 dimensions. Unfortunately, we do not know any $q=3$ tensor model and in order to somehow study this melonic model we shall consider a SYK like model with disorder, which is a special case of the models \cite{Murugan:2017eto}.
 
Thus, we shall try to study the following model
\begin{gather} \label{q3SYK}
S = \int d^d x\, d^2 \theta\,\left[\frac12 \left(D \Phi_i\right)^2 + C_{ijk} \Phi_{i}\Phi_{j}\Phi_{k} \right], \quad \braket{C^2_{ijk}} = \frac{J^2}{3 N^2}, i,j,k=1,\ldots,N,
\end{gather}
where we consider a quenched disorder for the coupling $C_{ijk}$. One might worry, that such a theory violates the causality, because the field $C_{ijk}$ is assumed to have the same value across the space-time and therefore the excitation of such a field changes the 
value of it everywhere, thus violating causality. But the procedure of quenching requires firstly to fix the value of $C_{ijk}$ that makes the theory casual and after that average over this field. It means that we can not excite the field $C_{ijk}$ and violate casuality.

\begin{figure}
	\centering
	\includegraphics[scale=0.64]{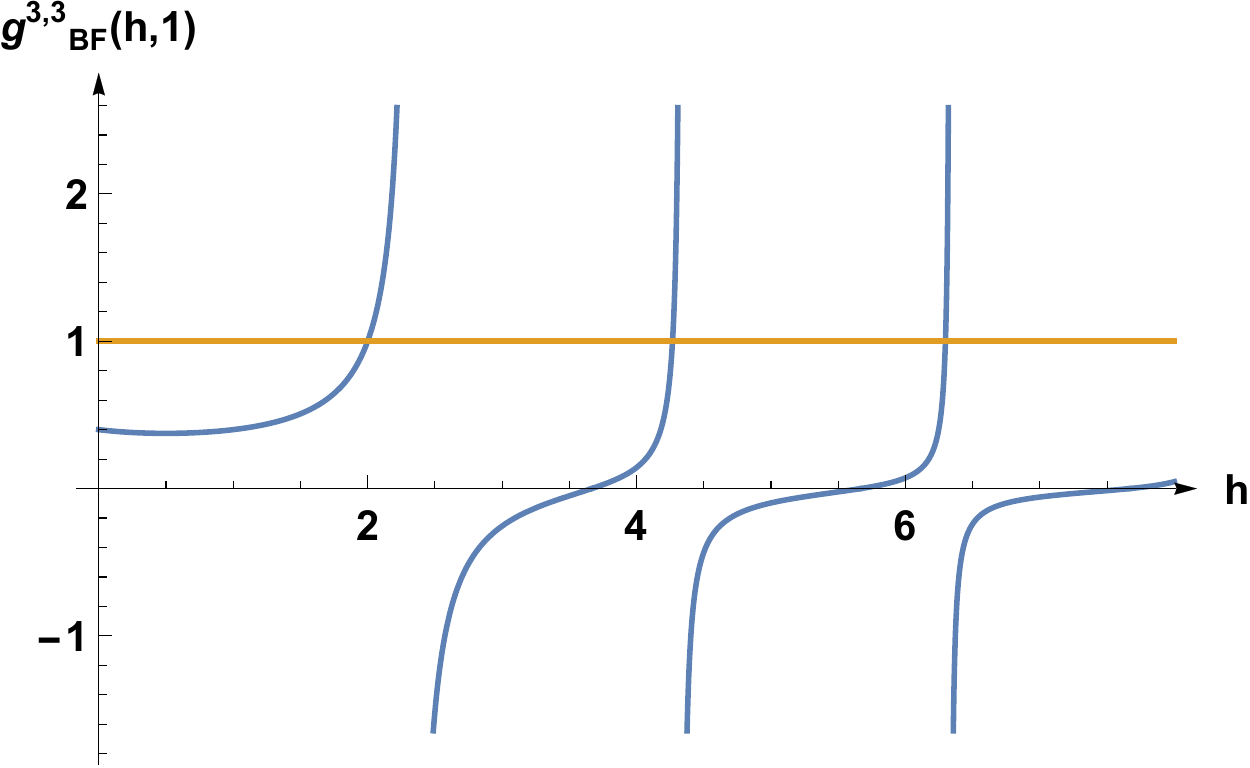}
	\includegraphics[scale=0.64]{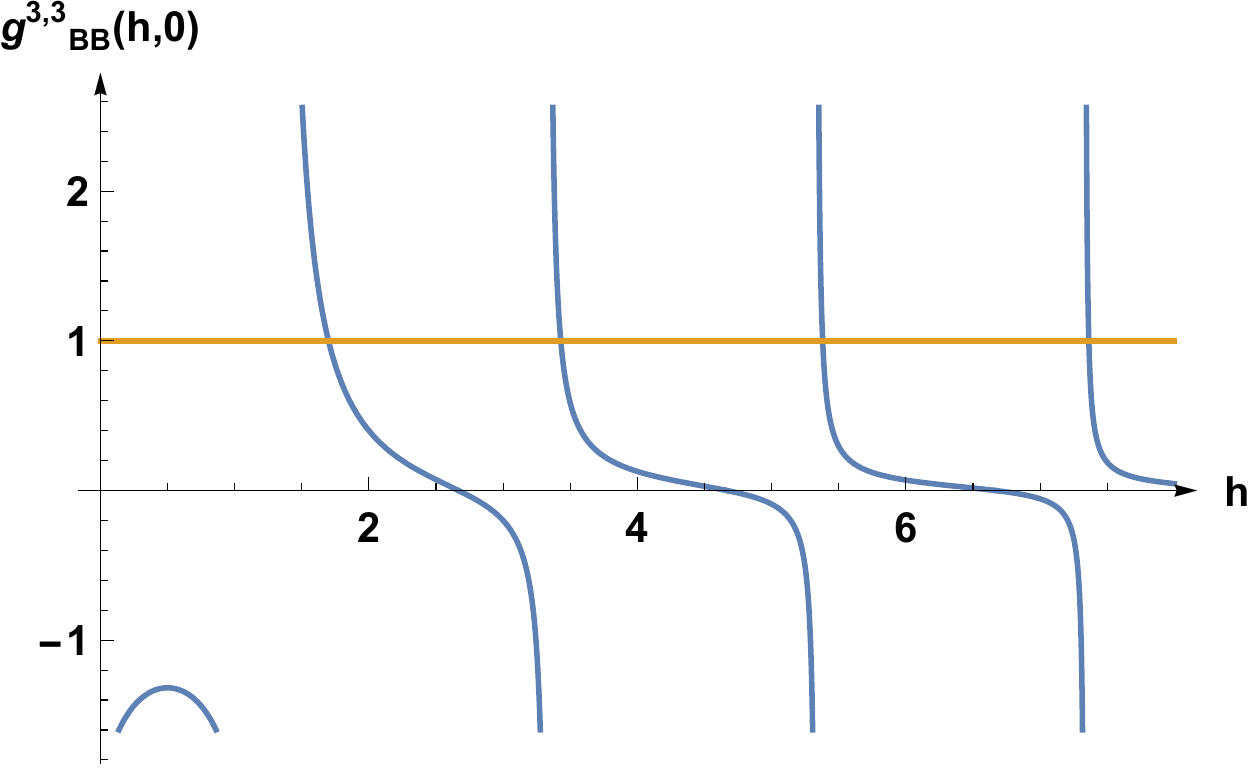}
	\caption{Plots for $g^{3,3}_{BF}(h,1)$ and $g^{3,3}_{BB}(h,0)$ that can help to understand the structure of the spectrum of the the theory \eqref{q3SYK}}
	\label{fig:2kh}
\end{figure}

This model is similar to the tensor one considered in the previous section, because again only melonic diagrams survive in the large $N$ limit, but with two iternal propagators in each melon. Therefore the formulas derived in the previous section are applicable in this case and with the replacement of $\lambda \to J$ and setting $q=3$ we can recover the large $N$ solution of this model.  For example, the propagator in this case is
\begin{gather}
G(x,\theta,\theta') = \frac{B}{\left|x_\mu - \bar{\theta}' \gamma_\mu \theta\right|^\frac43},\quad B^3 =\frac{1}{12\sqrt{3}\pi^3 J^2}, 
\end{gather}
and the dimension of the field $\Phi_i$ is $\Delta = \frac23$. Again the spectrum of the operators could be separated into three sectors, described in the previous section. The equation for the BB operators is determined by the equation
\begin{gather}
g^{3,3}_{BB}\left(h,s\right)= -\frac{2^\frac43 \sqrt{\pi}\Gamma\left(\frac23-\frac{h}{2}+\frac{s}{2}\right)\Gamma\left(\frac16 + \frac{h}{2}+\frac{s}{2}\right)}{3\Gamma\left(\frac16\right)\Gamma\left(\frac43-\frac{h}{2}+\frac{s}{2}\right)\Gamma\left(\frac56 + \frac{h}{2}+\frac{s}{2}\right)},
\end{gather}
where $s$ is the spin and should be chosen even. One can try to find the spectra of low lying states \eqref{fig:2kh}
\begin{center}
\begin{tabular}{|c|c|}
\hline
 $\left[\Phi^2\right]_{\theta=0} s=0$ & $\quad h=1.69944, 3.42951, 5.38013, 7.36259, 9.354,\ldots$ \\
\hline
 $\left[D_\alpha(\Phi^2)\right]_{\theta=0} s=1/2$ & $\quad h=2.19944, 3.92951, 5.88013, 7.86259, 9.854,\ldots$ \\
\hline
$\left[\Phi \partial_{\mu_1} \partial_{\mu_2} \Phi\right]_{\theta=0}  s=2$ &  $\quad h=3.51911, 5.39016, 7.3654, 9.35514, 11.3496, \ldots$ \\
\hline
$\left[D_\alpha(\Phi \partial_{\mu_1} \partial_{\mu_2} \Phi)\right]_{\theta=0} s=5/2$ & $\quad h=4.01911, 5.89016, 7.8654, 9.85514, 11.8496, \ldots$\\
\hline
\end{tabular}
\end{center}
It is easy to see that the spectrum has the following asymptotic behavior at large spins
\begin{equation*}
h\approx \frac{4}{3}+2n + s + \mathcal{O}\left(1/n,1/s\right),n\to \infty, s\to \infty.
\end{equation*}
On a principal line $h=\frac{d}{2}+i \alpha$ the kernel is complex, it is connected to the fact that there is no well-defined metric in the space of two-point functions \cite{Murugan:2017eto}. Therefore there is no problems with the complex modes, that could possibly destroy the conformal solution in the IR \cite{Kim:2019upg}. Thus $q=3$ supersymmetric SYK model is stable at least in the BB channel. Also one can check there are no additional solutions to the equation $g_{BB}^{3,3}(h,s)=1$ in the complex plane except the ones on the real line. The spectrum of the FF operators coincides with the spectrum of the BB operators but shifted with $h\to h+1$, therefore we don't have to worry about the instabilities of the theory in this sector.

Analogous calculations could be conducted for the BF series
\begin{gather}
g^{3,3}_{BF}\left(h,s\right) = -g^{3,3}_{BB}\left(h-\frac12,s-\frac12\right),
\end{gather}
where the spin $s$ should be in the form $s=2n-\frac12$.
One can notice that there is a solution $g^{3,3}_{BF}(5/2,3/2)=1$ corresponding to the existence of the supercurrent and energy momentum tensor (the energy momentum is not seen directly because it belongs to the supermultiplet of the supercurrent, but if one studies the theory in terms of the components, he or she will of course find the energy momentum tensor). There is a list of some low lying operators in the FF sector \eqref{fig:2kh}
\begin{center}
\begin{tabular}{|c|c|}
\hline
$\left[\partial_{\mu}\Phi D_\alpha \Phi\right]_{\theta=0}$ & 
$s=\frac32:  h=2.5,  4.76759,  6.79738, 8.80934, 10.8157, \ldots$\\
\hline
$\left[D_\beta\left(\partial_{\mu}\Phi D_\alpha \Phi\right)\right]_{\theta=0}$ & $s=2:  h=3,  5.26759,  7.29738, 9.30934, 11.3157, \ldots$\\
\hline
$\left[\partial_{\mu_1} \partial_{\mu_2} \partial_{\mu_3}\Phi D_\alpha \Phi\right]_{\theta=0}$ &
$s=\frac72: h=4.15398, 6.28752, 8.30627, 10.3143, 12.3189, \ldots$ \\
\hline
$\left[\partial_{\mu_1} \partial_{\mu_2} \partial_{\mu_3}D_\beta\left(\Phi D_\alpha \Phi\right)\right]_{\theta=0}$ & $s=4: h=4.65398, 6.78752, 8.80627, 10.8143, 12.8189, \ldots$\\
\hline
\end{tabular}
\end{center}
The spectrum has the following form asymptotic behavior
\begin{equation*}
h\approx \frac{5}{6}+2n + s + \mathcal{O}\left(1/n,1/s\right),n\to \infty.
\end{equation*}
The kernel is again complex on the principal line, but if one chooses $s=\frac12$ there would be an additional solution of the equation $g^{3,3}_{BF}=1$ at $h=1+0.496 i$, but as soon as it is not on the principal line and $s$ is not permissible we do not have to worry about this complex mode and expect that it could break the conformal solution. Thus this $q=3$ supersymmetric SYK model could provide us with a conformal field theory that is melonic and stable at integer dimensions. It would be interesting to study the $4-\eps$ expansion for this model, where it will be close to its critical dimension. 
\section{$3-\eps$ expansion}

In this section, we continue the investigation of the supersymmetric tensor model \eqref{smelon} from the point of view of the $\epsilon$ expansion. The calculation is similar to the ones performed in the papers \cite{Giombi:2017dtl,Giombi:2018qgp,Benedetti:2019eyl}. We include all possible $O(N)^3$ symmetric marginal interactions that respect the supersymmetry. Thus the superpotential has the following form
\begin{gather}
W(\Phi) = g_1 \Phi_{abc}\Phi_{ab'c'}\Phi_{a'bc'}\Phi_{a'b'c} + \notag\\+ \frac{g_2}{3}\left(\Phi_{abc}\Phi_{a'bc}\Phi_{ab'c'}\Phi_{a'b'c'}+\Phi_{abc}\Phi_{ab'c}\Phi_{a'bc'}\Phi_{a'b'c} + \Phi_{abc}\Phi_{abc'}\Phi_{a'b'c}\Phi_{a'b'c'}\right) + g_3 \left(\Phi_{abc}^2\right)^2,\label{gensuppot}
\end{gather}
where we imposed a symmetry under the exchange of the colors. In comparison to the "prismatic" theory \cite{Giombi:2018qgp}, which has $8$ coupling constants, the supersymmetric theory has only $3$; this is a significant simplification. 

Let us first consider the general renormalizable $d=3$ theory of ${\cal N}=1$ superfields $\Phi^i$, $i=1,\ldots n$:
\begin{gather}
S[\Phi_i] = \int d^3x d^2\theta\left[\frac12 (D\Phi_i)^2+ \frac{Y_{ijkl}}{4!} \Phi_i\Phi_j\Phi_k\Phi_l\right],
\end{gather}
where $Y_{ijkl}$ is a real symmetric tensor. Adapting the results from \cite{avdeev1992renormalizations,Jack:2015tka}, we find that
the two-loop corrections to the gamma and beta functions are
\begin{gather} 
\gamma^{(2)}_{ab} = \frac{1}{3(8\pi)^2} Y_{ajkl}Y_{bjkl},\notag\\
\beta^{(2)}_{abcd} = \frac{1}{3(8\pi)^2} Y_{ijkl}\left (Y_{jkla} Y_{bcdi}+Y_{jklb} Y_{cdai}+Y_{jklc} Y_{acdi}+Y_{jkld} Y_{abci}\right )+\notag\\+\frac{2}{(8\pi)^2}\left (Y_{anom}Y_{bfom}Y_{nfcd}+Y_{anom}Y_{cfom}Y_{nfbd}+Y_{anom}Y_{dfom}Y_{nfbc}+\right.\notag\\\left.+Y_{bnom}Y_{cfom}Y_{nfad}+Y_{bnom}Y_{dfom}Y_{nfac}+Y_{cnom}Y_{dfom}Y_{nfab}\right ) . \label{ANS}
\end{gather}
These two-loop results are closely related to those in a non-supersymmetric theory with Yukawa coupling 
$\frac{1}{4} Y_{ijkl} \psi^i \psi^j \phi^k \phi^l$ (see \cite{Jack:2015tka}), except the supersymmetry requires  $Y_{ijkl}$ to be fully symmetric.

Substituting $Y_{ijkl}$ corresponding to the superpotential (\ref{gensuppot}), we find from \eqref{ANS}:
\begin{gather}
\gamma^\Phi_{abc,a'b'c'} = \delta_{aa'}\delta_{bb'}\delta_{cc'}\gamma^\Phi\notag\\
\gamma^\Phi=\frac{1}{6\pi^2}\left[12 g_2 g_1(1+N+N^2)+6 g_3^2(2+N^3) + 3 g_1^2(2+3N+N^3)+ \right.\notag\\\left.+ g_2^2(5+9N+3N^2+N^3)+ 36 g_3 g_1 N + 12 g_3 g_2(1+N+N^2)\right]
\end{gather}
and
\begin{gather}\label{sbetaexp}
\beta_1 =-\eps g_1+ \frac{2}{9 \pi ^2} \left(6 g_1 \left(12 g_3^2 \left(N^3+11\right)+g_2^2 (N^3+6 N^2+30 N+29)+12 g_3 g_2 (2 N^2 + 5 N +5) \right)+\right.\notag\\ \left. +9 g_1^3 \left(N^3+12 N+8\right)+18 g_1^2 \left(g_2 (4 N^2+7 N+16)+24 g_3 N\right)+2 g_2^2 \left(g_2 (2 N^2 + 13 N+24)+72
	g_3\right)\right),\notag\\
\beta_2 = - \eps g_2+ \frac{2}{9 \pi ^2} \left(g_2 \left(72 g_3^2 \left(N^3+11\right)+g_2^2 (7 N^3+36N^2+162 N+194)+36 g_3 g_2 ( (5 N^2 +9N +16)\right)+\right.\notag\\ \left. +54 g_1^3 \left(N^2+N+4\right)+18 g_1^2 \left(g_2 (N^3+3 N^2+27N+26)+18 g_3 (N+2)\right)+ \right.\notag\\ \left. + 18 g_2 g_1 \left(g_2 (7 N^2+21 N+32)+48 g_3
	(N+1)\right)\right),\notag\\
\beta_3 = -\eps g_3 + \frac{2}{9 \pi ^2} \left(108 g_3^3 \left(N^3+4\right)+252 g_2 g_3^2 \left(N^2+N+1\right)+7 g_2^3 (N^2+3 N+5)+ \right.\notag\\\left.+ 18 g_1^2 \left(2 g_3 \left(N^3+3 N+2\right)+g_2 \left(N^2+N+4\right)\right)+27 g_1^3 N++12 g_2^2 g_3
(N^3+3N^2+15N+14)+\right.\notag\\ \left. 36 g_1 \left(2 g_2^2 (N+1)+2 g_3 g_2 (2 N^2 +2 N+5)+21 g_3^2 N\right)\right)
\end{gather}
If one sets $g_1=g_2=0$, the symmetry gets enhanced to $O(N^3)$ and corresponds to the $O(n)$ vector model, which was considered in \cite{avdeev1992renormalizations}.\footnote{Please note that they considered $SU(n)$ case that corresponds to $N^3=2n$ and their definition of $\gamma^\Phi$ includes a factor of two.} 
For the supersymmetric $O(n)$ model with superpotential $g (\Phi^i \Phi^i)^2$,
\begin{equation}
\beta_g =- \eps g + \frac{24 (n+4)}{\pi^2} g^3 + O(g^5)\ ,
\end{equation}
in agreement with \cite{avdeev1992renormalizations}.

If we choose $N=1$, the couplings $g_1,g_2,g_3$ becomes degenerate because they describe the same operator. Therefore, the beta-functions should be added to get the right expression. And indeed, if we choose $N=1$ and sum up the couplings we get
\begin{gather}
\beta_1+\beta_2+\beta_3 =\mu \frac{d(g_1+g_2+g_3)}{d\mu} =  - \eps(g_1+g_2+g_3) + \frac{120}{\pi^2}(g_1+g_2+g_3)^3,
\end{gather}
which is the correct beta function for the theory with superpotential $(g_1+ g_2+ g_3) \Phi^4$ for a single chiral superfield $\Phi$.
This special case of our theory is conformal in the entire range $2\leq d < 3$. Indeed, in $d=2$ the ${\cal N}=1$ supersymmetric theory with superpotential $\Phi^m$ for one superfield $\Phi$ flows to the
superconformal minimal model with central charge 
\begin{equation}
c= \frac{3} {2} \left ( 1- \frac{8}{m(m+2)} \right )\ .
\end{equation}
Therefore, the $N=1$ case of the supertensor model gives the $m=4$, $c=1$ superminimal model in $d=2$. For $N>2$ the $O(N)^3$ supertensor model is expected to be conformal
in $2<d<3$, but not in $d=2$. 

Let us consider the large $N$ limit where we scale the coupling constants in the following way:
\begin{gather}\label{scaling}
g_1 = \frac{\pi}{2}\frac{\sqrt{2\eps}\lambda_1}{N^\frac32},\quad g_2 = \frac{\pi}{2}\frac{\sqrt{2\eps}\lambda_2}{N^\frac52},\quad g_3 = \frac{\pi}{2}\frac{\sqrt{2\eps}\lambda_3}{N^\frac72}.
\end{gather}
The scaling is taken to be the same as in the 
paper \cite{Giombi:2017dtl}.
Applying this scaling to the formula \eqref{sbetaexp}, we get 
\begin{gather}
\gamma_\Phi = \eps\frac{\lambda_1^2}{4}, \qquad \beta_1 = - \lambda_1 + \lambda_1^3, \\
\beta_2 = -  \lambda_2+ 2\lambda_2 \lambda_1^2+6\lambda_1^3, \qquad \beta_3 = -\lambda_3 + 2(2\lambda_3+\lambda_2) \lambda_1^2 + 3\lambda_1^3\ .\notag
\end{gather}
From this one can find the fixed point in the large $N$ limit. Namely,
\begin{gather}\label{fixpoint}
\lambda^\infty_1 = \pm 1,\quad \lambda^\infty_2 =\mp 6,\quad \lambda^\infty_{3} = \pm 3,\quad \Delta_\Phi = \frac{d-2}{2}+\gamma_\Phi=\frac12-\frac{\eps}{4}.
\end{gather}
We may  try to compute the $1/N$ corrections to these results to get
\begin{gather}
\lambda_1 = 1 + \mathcal{O}\left(\frac{1}{N^2}\right),\quad \lambda_2 = -6 + \frac{20}{N}+ \mathcal{O}\left(\frac{1}{N^2}\right),\notag\\ \lambda_{3} = 3 - \frac{16}{N}+ \mathcal{O}\left(\frac{1}{N^2}\right),\quad \gamma^\Phi = \frac12- \frac{\eps}{4}+\mathcal{O}\left(\frac{1}{N^2}\right).
\end{gather}
The anomalous dimension of the matter field operator $\Phi$ coincides with the  exact dimension of the field by solving the DS equation found above. This might indicate that the higher-loop corrections to the RG equations \eqref{sbetaexp} are suppressed in the large $N$ limit. It would be interesting to study these suppressions in $N$ for a general superpotential \eqref{gensuppot} from a combinatorial diagrammatic point of view and compare the results with the investigation of the finite $N$ solutions of the equations \eqref{sbetaexp}.

If one considers the large $N$ fixed point \eqref{fixpoint} of the RG flow governed by the equations \eqref{sbetaexp} and tries to descend to finite $N$, one can find that  the solution always exists (see the table \eqref{thetable}) and quite close to the found fixed point \eqref{fixpoint} (of course with the appropriate chosen scaling), in comparison to the "prismatic" model, where the melonic fixed point exists only at $N>54$ \cite{Giombi:2018qgp}.
\begin{table}
		\centering
	\begin{tabular}{ |c|c|c|c| } 
		\hline
		$N$ & $\frac{\lambda_1}{\lambda_1^\infty}$ & $\frac{\lambda_2}{\lambda_2^\infty}$ & $\frac{\lambda_3}{\lambda_3^\infty}$ \\ 
		\hline
		100000 & 1.000 & 1.000 & 1.000\\ 
		\hline
		10000 &  1.000 & 1.001 & 1.002 \\ 
		\hline
		1000 & 1.000 & 0.995 &  0.995\\
		\hline
		100 & 1.001 & 0.953 & 0.950\\
		\hline
		10 & 1.033 & 0.691 & 0.670\\
		\hline
		5 & 1.068 & 0.546 & 0.527\\
		\hline
		2 & 1.049 & 0.350 & 0.322\\
		\hline
		1 & 1.093 & 0.273& 0.139\\
		\hline
	\end{tabular}
	\caption{The approach of the finite $N$ fixed points in $3-\epsilon$ dimensions to 
		the large $N$ limit. We note that the fixed point exists for all values of 
		$N$.
	}
	\label{thetable}
\end{table}

We can study the dimension of various operators in the fixed point \eqref{fixpoint}. One of these operators is $\Phi_{abc}^2$, which belongs to the $BB$ spectrum. We can find that the anomalous dimension of this operator is
\begin{gather}
\Delta_{\Phi^2} = \Delta^0_{\Phi^2} + 2\gamma_\Phi+\gamma_{\Phi^2}= 1 + \eps+ \mathcal{O}(\eps^2),
\end{gather}
where we have used the relation $\gamma_{\Phi^2} = 6 \gamma_\Phi$, which is true only at the second level of perturbation theory. The answer coincides with the exact solution found earlier \eqref{dimPhi2}.

As one can see, the fixed point \eqref{fixpoint} is IR stable, which means that the dimensions of the operators is bigger than the dimension of the space-time. Indeed, the linearized equations of RG flow near the fixed point \eqref{fixpoint} have the following eigenvalues
\begin{gather}
\left(\frac{\partial \beta_i}{\partial \lambda_j}\right) = \begin{pmatrix} -1+ 3\lambda_1^2 & 0 & 0\\ 4\lambda_2 \lambda_1 + 18\lambda_1^2 & -1+2\lambda_1^2  & 0 \\  4(2\lambda_3+\lambda_2)\lambda_1+9\lambda_1^2 &   2\lambda_1^2 & -1 + 4\lambda_1^2
\end{pmatrix},\quad \Lambda = \left[2,1,3\right],
\end{gather}but as it is known the eigenvalues of this matrix gives the dimensions of quartic operators
\begin{gather}
\Delta_i=d-\eps +\Lambda_i.
\end{gather}
Thus we get
\begin{gather}
\Delta_{\Phi^4} = 2 - \eps + 3 \eps = 2 + 2 \eps + \mathcal{O}(\eps^2),\quad
\Delta_{\rm pillow} = 2 - \eps + \eps = 2 + \mathcal{O}(\eps^2),\notag\\
\Delta_{\rm tetra} = 2 -  \eps+ 2 \eps = 2 + \eps + \mathcal{O}(\eps^2). 
\end{gather}
This is in the agreement with the large $N$ solution.
As one can see, $\Lambda_i>0$, indicating that the fixed point is IR stable. The agreement found between the exact large $N$ solution and perturbative $\eps$ expansion indicates that there is a nice flow from the UV scale to the IR one where the bare, free propagator flows to the one found by direct solving the DS equations \eqref{DSeq}. The study of the higher loop corrections might help to understand this relation better.
\section{$\N=2$ supersymmetry and gauging}
One can try to consider $\N=2$ supersymmetry and study the properties of such a model. Here we are not going to present the solution of the corresponding DS equation , but we will just calculate the beta-functions and find the fixed point of the resulting equations. The SYK model with $\N=2$ supersymmetry at $2$ dimensions was considered in the paper \cite{Bulycheva:2018qcp}.

The theory is built analogously to the $\N=1$ case. It can be obtained by dimensional reduction from $\N=1$ supersymmetry in 4 dimensions. In this case, we have a set of chiral superfields $\Psi_{abc}$ with the action
\begin{gather}
S = \int d^3 x \,d^2\theta\, d^2\bar{\theta}\, \bar{\Psi}_{abc}\Psi_{abc} + \int d^3 x\, d^2\theta\, W(\Psi_{abc}) + {\rm h.c.},\quad \bar{D}_\alpha \Psi_{abc}=0,
\end{gather}
where the superpotential is taken to be the same as in the case of $\N=1$ supersymmetry. The beta-function for a general quartic superpotential was considered in the paper \cite{gracey2017function}. The beta-function receives corrections only from the field renormalizations, meaning that it has the following form
\begin{gather}
\beta_{1,2,3} = \left(-\eps+4\gamma^\Phi\right)g_{1,2,3} \notag\\
\gamma^\Phi = \frac{1}{6\pi^2}\left(12 g_2 g_1\left(1+N+N^2\right) + 6 g_3^2 (2+N^3) + 3 g_1^2 (2+3N+N^3) +\right.\notag\\\left.+ g_2^2 (5+9N+3N^2 + N^3) + 36 g_3 g_1 N + 12 g_3 g_2(1+N+N^2)\right) \label{gammaN2}.
\end{gather}
The fixed point is determined by demanding that the anomalous dimension of the field must be $\Delta_\Phi=\Delta^0_\Phi+\gamma^\Phi=\frac{d-1}{4}$, as we got for a general melonic theory in arbitrary dimensions. Apparently, for $\N=2$ models this fact comes not from the melonic dominance, but from the consideration of the supersymmetric algebra that fixes the dimensions to be proportional to the $R$ charge of the corresponding operator. This condition defines a whole manifold in the space of marginal couplings. Applying the scaling \eqref{scaling}, in the large $N$ limit we get the equation
\begin{gather}
\gamma(\lambda_1,\lambda_2,\lambda_3) = \frac{\lambda_1^2}{4} = \frac{1}{4}, \quad \lambda_1 =1. 
\end{gather}
It is quite interesting that this equation does not fix $\lambda_2,\lambda_3$ in the large $N$ limit. One can study the stability of these fixed points at arbitrary $\lambda_{2,3}$. The RG flow near the fixed point could be linearized to get the stability matrix
\begin{gather}
\left(\frac{\partial \beta_i}{\partial g_j}\right) = \begin{pmatrix}
2 & 0 & 0\\
2\lambda_2 & 0 &  0\\
2\lambda_2 & 0 & 0
\end{pmatrix}, \quad \Lambda = \left[2,0,0\right].
\end{gather}
The given solution is marginally stable, because of the existence of two marginal operators. These two zero directions correspond to the previously discussed existence of a whole manifold of IR fixed points. 

From this consideration, it would be interesting to study the large $N$ limit of the considered $\N=2$ theory and corresponding DS equations. This model must have the same combinatorial properties as the $\N=1$ and scalar tensor model, but some cancellation happens that drastically simplifies the theory.

One can try to examine a gauged version of $\N=2$ theory. The gauging of the tensor models is one of the important aspects that makes them different from the SYK model. In the latter, due to the presence of the disorder in the system, the theory can possess only the global $O(N)$ symmetry and can not be gauged, while in the tensor models there are no such obstructions and one can add gauge field and couple to the tensor models at any dimensions. 

Gauging should be important for understanding the actual AdS/CFT correspondence. In 1 dimension, the gauging singles out from the spectrum all non-singlet states from the Hilbert states. There have been many attempts to understand of the structure of the tensorial quantum mechanics of Majorana fermions from numerical and analytical calculations \cite{Klebanov:2018nfp,Pakrouski:2018jcc, Krishnan:2017txw, Krishnan:2018hhu}. These gave some interesting results, such as the structure of the spectrum of the matrix quantum mechanics and the importance of the discrete symmetries for explaining huge degenaracies of the spectra. Still, the general impact of gauging of the tensorial theory is not clear and demands a new approach. Here, we will give some comments of the combinatorial character and study how the gauging of $\N=2$ theory, studied in the previous section, changes.

In 3 dimensions one can gauge a theory by adding a Chern-Simons term instead of the usual Yang-Mills term
\begin{gather}
S= \int d^3 x d^2\theta\left[-k\left(D_\alpha\Gamma^a_\beta\right)^2 + \left|\left(D_\alpha \delta^a_b  +g \Gamma^a_{b\,\alpha} \right) T^a\Phi_{abc}\right|^2 + W(\Phi_{abc})\right],
\end{gather}
where $W(\Phi_{abc})$ is the same as in the \eqref{gensuppot}, $T^a$ are the generators of the group $O(N)\times O(N)\times O(N)$, and $\Gamma^\alpha$ are vector superfields that have a  gauge potential $A^a_{b\mu}$ as one of the components. If one rewrites the kinetic term for the  gauge  field in terms of usual components, he will get a usual Chern-Simons theory.  Since the theory is gauge invariant, we can choose an axial gauge to simplifty the action \footnote{I would like to thank S.Prakash for the suggested argument.} $A^{a}_{b3}=0$, which eliminates the non-linear term from the theory and the Fadeev-Popov ghosts decouple from the theory. Therefore the $A^{a}_{b1}, A^{a}_{b2}$ can be integrated out to get an effective potential. For example, such a term appears in the action
\begin{gather}\label{effectivesuper}
W_{\rm eff} \sim \frac{1}{k}\int \frac{d^3 q}{(2\pi)^3}\frac{(\Phi_{abc} D_\alpha\Phi_{ab'c'})(q)(\Phi_{a'bc}D_\alpha\Phi_{a'b'c'})(-q)}{q_\perp} + \text{perm.},
\end{gather}
which can be considered as a non-local pillow operator with the wrong scaling, because the level of CS action usually scales as $k=\lambda N$.  Therefore some diagrams would have large $N$ factor and diverge in the large $N$ limit. To fix it we should consider the unusual scaling for the CS level $k=\lambda N^2$. 

One can check that only specific Feynman propagators containing the non-local vertex \eqref{effectivesuper} contribute in the large $N$ limit \cite{Benedetti:2019eyl}. Namely only snail diagrams contribute in the large $N$ limit and usually are equal to zero by dimensional regularization for massless fields. Therefore, one can suggest that the gauge field in the large $N$ limit does not get any large corrections and does not change the dynamics of the theory. This argument being purely combinatorial should be applied for any theory coupled to the CS action.

We can confirm this argument by direct calculation of the dimensions of the fields in the $\epsilon$ expansion for the $\N=2$ supertensor model at two-loops and see whether the dimensions of the fields gets modified.  The beta-functions for a general $\N=2$ theory coupled to a CS action was considered in the paper \cite{gracey2017function} and have the following form at finite $N$
\begin{gather}
\beta_{1,2,3} = \left(-\eps + 4\gamma^\Phi_k\right) g_{1,2,3},\quad
\gamma^\Phi_k = \gamma^\Phi - \frac{3 N(N-1)}{64\pi^2 k^2},
\end{gather} 
where $\gamma^\Phi$ is the same as in the equation \eqref{gammaN2}.
As $k\sim N^2, N\to \infty$ the corrections to the gamma-functions vanish in the large $N$ limit. Thus, the gauging in three dimensions indeed does not bring any new corrections to the theory. It would be interesting to study such a behavior in different dimensions. For example, if in 1 dimension the gauging does not change structure of the solutions,  one may conclude that the main physical degrees of freedom are singlets and there is a gap between the non-singlet and singlet sectors. Also it would be interesting to confirm this observation by a direct computation for the prismatic theories and for Yang-Mills theories.

\section*{Acknowledgments}
I would like to thank Igor R. Klebanov for suggesting this  problem and for the guidance throughout the project.
This research was supported in part by the US NSF under Grant No.~PHY-1620059.  I am very grateful to S. Giombi and G. Tarnopolsky for collaboration at the early stages of this project.
The author also thank  A.M.Polyakov, S.Prakash, E. Akhmedov and A.Milekhin   for useful discussions.  Also I would like to thank M.Grinberg and P.Pallegar for the careful reading of the first drafts. I thank the organizers of the conference ``Quantum Gravity 2019" in Paris
for hospitality and stimulating atmosphere during some of the work of this project. 

I dedicate this work to the memory of my physics teacher, Polyanskii Sergey Evgenievich. Sergey Evgenievich guided me throughout School No.146 in Perm, Moscow Insitute of Physics and Technology and shared his profound wisdom, that helped me to come to the point where I am. 
 
\appendix
\section{Supersymmetry in 3 dimensions}
 \label{supersection}
In this section we will introduce the notations and useful identities for the ${\cal N}=1$ supersymmetric theories in 3 dimensions. We will mostly follow the lectures \cite{gates2001superspace}. The Lorentz group in 3 dimensions is $SL(2,\mathbb{R})$; that is a group of all unimodular real matrices of dimension 2. The gamma matrices can be chosen to be real
\begin{gather}
\gamma^0 = \begin{pmatrix}
0 & -1\\
1 & 0
\end{pmatrix},\quad \gamma^1 = \begin{pmatrix}
0 & 1\\
1 & 0
\end{pmatrix},\quad \gamma^2 = \begin{pmatrix}
1 & 0\\
0 & -1
\end{pmatrix}, \quad \left\{\gamma^\mu,\gamma^\nu \right\} = 2 \eta^{\mu\nu}.
\end{gather}
There is no $\gamma^5$ matrix, so we can't split the spinor representation into small Weyl ones. Because of this, the smallest spinor representation is 2 dimensional and real. It is endowed with a scalar product defined as
\begin{gather}
\bar{\xi}\eta = \xi^\alpha\eta_\alpha= i\xi^\alpha \gamma^0_{\alpha\beta} \eta^\beta,\quad \theta^2 =\frac12 \bar{\theta}\theta.
\end{gather}
Because of these facts, the $\N=1$ superspace, in addition to the usual space-time coordinates, will include two real Grassman variables $\theta^\pm$.
The fields on the superspace can be decomposed in terms of fields in the usual Minkowski space. For instance, a scalar superfield (that is our major interest) has the following decomposition
\begin{gather}
\Phi(x,\theta^\alpha) = \phi(x) + \bar{\theta}\psi(x) + \theta^2  F(x). \label{superfield}
\end{gather}
As usual, the algebra supersymmetry in superspace can be realized via the derivatives that act on the superfields \eqref{superfield} and mix different components
\begin{gather}
Q_\alpha = \partial_\alpha + i\gamma^\mu_{\alpha\beta}\theta^\beta\partial_\mu,\quad \left\{Q_\alpha,Q_\beta \right\} = 2 i \gamma^\mu_{\alpha\beta}\partial_\mu
\end{gather} 
where $\partial_\mu$ stands for differentiation with respect to the usual space-time variables, and $\partial_\alpha$ for the anticommuting ones. One can define a superderivative that anticommutes with supersymmetry generators, and therefore preserves the supersymmetry
\begin{gather}
\label{sderv}
D_\alpha = \partial_\alpha - i \gamma^\mu_{\alpha\beta}\theta^\beta \partial_\mu,\quad \left\{D_\alpha, Q_\beta \right\}  = 0.
\end{gather}
Out of these ingredients, namely \eqref{superfield},\eqref{sderv}, we can build an explicit version of a supersymmetric Lagrangian. For example, we can consider the following Lagrangian
\begin{gather}\label{slag}
S = \int d^3 x d^2 \theta\left[-\frac12 \left(D_\alpha \Phi\right)^2 + W(\Phi)\right],
\end{gather}
where the integral over Grassman variables is defined in the usual way with the normalization $\int d^2 \theta \bar{\theta}\theta = 1$. Writing out the explicit form of \eqref{slag} we get
\begin{gather}
S = \int d^3 x \left[\frac12(\partial_\mu \phi)^2 + i \psi^{\alpha}\gamma^\mu_{\alpha\beta}\partial_\mu \psi^\beta + F^2+W'(\phi) F + W''(\phi)\psi^2\right].
\end{gather}
The field $F$ does not have a kinetic term, and therefore is not dynamical and can be integrated out (that we will not do). For a further investigation we have to develop the technique of super Feynman graphs. We start with considering the partition function of the theory \eqref{slag}
\begin{gather}
Z[J] = \int [d\Phi] \exp\left[\int d^3 x d^2\theta \left(\frac12 \left(D_\alpha \Phi\right)^2 + W(\Phi) + J \Phi \right)\right] =\notag\\ =  \exp\left(W\left(\frac{\delta}{\delta J}\right)\right) \int [d\Phi] \exp\left[\int d^3 x d^2\theta \left(\frac12 \Phi D^2 \Phi + J \Phi \right)\right].  \label{spart}
\end{gather} 
The last integral is gaussian and therefore can be evaluated and is equal to
\begin{gather}
Z[J] = \exp\left(W\left(\frac{\delta}{\delta J}\right)\right) \exp\left(-\int d^3 x d^2 \theta \left[ \frac12 J \frac{1}{D^2} J\right]\right).
\end{gather}
From this one can recover the usual Feynman diagrammatic technique, where the vertex is taken from the superpotential $W(\Phi)$ rather than the integrated version, and the propagator is defined as
\begin{gather}
\braket{\Phi(x_1,\theta_1)\Phi(x_2,\theta_2)} = \frac{1}{D^2} \delta^2(\theta_1-\theta_2) = \frac{D^2}{\Box} \delta^2(\theta_1-\theta_2), \label{sprop}
\end{gather}
which can be calculated by double differentiation of the partition function \eqref{spart},
and the operator $\Box$ is the usual laplacian.
\bibliographystyle{ssg}
\bibliography{STM}

\end{document}